# Object-Based Groupware: Theory, Design and Implementation Issues
## (Extended Absract)

**Sergey V. Zykov** (sz@aha.ru)
**Gleb G. Pogodayev** (skiba@cv.jinr.dubna.ru)
Kashirskoye Avenue, 31
Cybernetics Department
Moscow Engineering Physics Institute
Moscow, 115409, Russian Federation

**Abstract**
Document management software systems are having a wide audience at present. However, groupware as a term has a wide variety of possible definitions. Groupware classification attempt is made in this paper. Possible approaches to groupware are considered including document management, document control and mailing systems. Lattice theory and concept modelling are presented as a theoretical background for the systems in question. Current technologies in state-of-the-art document mangenent software are discussed. Design and implementation aspects for user-friendly integrate enterprise systems are described. Results for a real system to be implemented are given. Perspectives of the field in question are discussed.

**1. Introduction**
Groupware is a collection of technologies that allow us to represent complex processes that center around collaborative human activities. It is based on the following technologies:
- multimedia document management;
- workflow;
- e-mail;
- conferencing;
- scheduling.

The interest in groupware is constantly arising since it essentially transforms many kinds of communication activities peole are engaged in companies. According to the Workgroup Technologies market research groupware sales are supposed to grow more than 2.5 times in 1998 compared to 1995 level. Annual grouth rate of the top groupware product - Lotus Notes - exceeds 50% [7].Groupware helps to remove layers of bureaucracy by optimising documents storage, retrieval and exchange. It also helps to regulate and order workflows and make collaborative work more efficient during each and every stage of projects lifecycle.
Many groupware instances allow to develop and deploy their own applications which results in substantial returns on investment.

**2. Existing Approaches and Technologies**

Current approaches to groupware are based on specific software accents existing at present.
Document management software can be roughly didvided into three major classes:
- document control systems (including workflow and scheduling technologies);
- document management systems (including multimedia management technologies);
- document mail systems (including e-mail and conferencing technologies).

The first and the third ones are designed to manage document flow while the second one is made for documents archiving and maintanance. Let us briefly describe each of the approaches named.

**2.1. Basic Document Management Approach**
includes the following services:
- search and access service;
- storage service;
- security service;
- version control service;
- archival and deletion service;
- integration service.

Search and access service enables users to quickly and effectively access the information they are searching for depending on the rights granted to the user group and access level specified. Storage service provides mechanisms for automatic storing documents on the network server. Security service grants document-level and network-level access control. The system also features an optional security server module. Version control service keeps track on documents corrected frequently by variuos users. Archival and deletion service enables system administrator to select the documents which do not need on-line access and to move them to a different storage or to delete them. Integration service allows the document management system not to exist separately from document generating applications (e.g. word processors, spreadsheets etc.) but rather to be interrelated with these applications.

A typical example of document management software is Novell SoftSolutions.

**2.2. Profile (Form) Control Approach**
usually features
- form designer;

- financial functions calculator;
- data validation checker;
- major SQL database formats support;
- ODBC

Delrina FormFlow is profile (or form) control software instance[7].

## 2.3. Multi-Format Document Mailing Approach

features the following major services:
- integrated message exchange service;
- multy-platform support;
- various connection type options;
- mobile user support;
- administrator tools.

Typical multi-format document mailing systems are Novell GroupWise and its successor - GroupWise XTD [7].

## 2.4. Integrated approach

There are, however, even more versatile approaches existing which are a combination of two or more of mentioned ones. The two most successful examples include Oracle InterOffice [15]and Lotus Notes [7].

### 2.4.1. Oracle InterOffice

Collaborative services of Oracle InterOffice are based on scalable and reliable relational database Oracle server and include World Wide Web facilities as well.

Oracle InterOffice built on a client/server architecture includes the following collaborative functions:
- comprehensive messagimg;
- directory services;
- calendar and resource scheduling;
- discussion;
- document management;
- workflow facilities.

### 2.4.1. Lotus Notes

Lotus Notes allows groups of users to interact and share information that be a highly unstructured nature. Its client/server application develop run-time environment provide the following functions.
- A document database server stores and manages multiuser client at semi-structured data-including text, images, audio, and video.Release 4 supports up to 1000 concurrent Notes clients using 32-bit SMP server platforms.
- An e-mail server manages multiuser client access to mail. Notes ñomes with a mail backbone infrastructure; X.400 is available as an optional component. Release 4 fully integrates X.400 and Internet SMTP/MIME backbone.
- A backbone server/server infrastructure supports both mail-route database replication. The replication mechanism synchronizes copies of the same database, which can reside on multiple server (or client) machines. Release 4 now supports a new server passthrough feature; it lets you d one Notes server and reach any other server to which you are auth Passthrough also lets you access multiple databases on multiple servers same time.
- A GUI client environment presents views of the document databases and provides an electronic mail front-end. Users can navigate through the databases and their document contents. Views are stored queries that act as filters for the information in the databases. The e-mail front-end is just a specialized view of a mail database. Notes can attach GUI forms (private or public) to the various databases used for data entry. Release 4 introduced the new Notes Mail client; it supports a flexible three-pane user interface that integrates cc:Mail, Web browsing, and the traditional Notes client. It also provides a universal in-box that can accept all types of mail, faxes, and forms.
- Distributed services include electronic signatures, security and access control lists, database administration services, system management, and an X.500 based global namespace.
- Application development tools include: 1) a GUI forms generator; 2) tools and templates for creating databases; 3) a primitive scripting language consisting of formulas, and 4) an open API set including the Notes API, VIM, MAPI, and ODBC. Release 4 supports Intelligent Agents, These are scripted Notes applications that you can use to automate repetitive tasks - including database replication, data-handling, the in-box actions, and messaging services. Release 4 also introduces LotusScript 3.0 - a cross-platform, BASIC-like, object-oriented programming language. In addition, you can write Notes applications using many popular third-party client/server tools - including Delphi, SQLWindows, PowerBuilder, New Era, VisualAge, Notes ViP, and Visual Basic (via the Lotus-provided HiTest VBXs).

## 3. References

[1-4, 6, 9] provide rigorous mathematics foundation and solid theoretical background for the paper.

Lattice of flow diagarams which can be used to model workflow are presented in [1].

ISA hierarchy as a basic appproach to objects storage and manipulation handling is describeed in [4]

Papers [3,4,9] deal with various database notations and help to compare relational, hierarchical and object databases parameters. [3] contains a detailed description an detailed description of relational database theoretical background versus that for network one in [4].
Various aspects of object-based implementations are discussed in [9].

The paper [5] deals with structured data representations. It is stated that there is no single tool at present able to deal with linear, table and hierarchical document structures.

Client/server aspects of groupware are presented in [7]. Groupware classification is given in the paper. Major vendors products are compared.

Workflow as an essential groupware technology is discussed in [8]. Outlines for enterprise solutions are given.

Information on major vendors of current groupware is presented both in hot periodicals [10-12] and WWW sites [13-15].

## 4. Design Suggestions

When designing any office system and an enterprise system in particular it is recommended to combine
- CASE or RAD tools or similar technologies;
- advanced DB systems.

The following features are highly recommended to design state-of-the-art enterprise document management, control and mailing systems:
- scalability;
- distributed multi-platform database processing;
- spatial data management;
- advanced version control facilities;
- multimedia data handling;
- advanced transaction management;
- query optimization facilities;
- archiving;
- Internet and Intranet access.

## 5. Systems in Question and Object Theories
In case of document management, control and mailing systems the following basic operations are necessary:
- documents (information) flow control;
- documents structuring and archival;
- documents search and updating.

This basically means in terms of conceptual modelling that a model of the future system is to be built.

To build a system in general it is necessary to perform the following actions:
- define document routes;
- define rules according to which documents are stored and processed;
- define roles of objects (people processing documents etc.)

Let us consider a fairly large enterprise with respective amount of documents processed (hundreds of them daily). Authors in this paper tend to describe some basic procedures and common laws rather than explicitly define each and every document type and possible route.

### 5.1. Document Hierarchy
Some basic examples of document types can be summarised as follows:
- order;
- decree;
- Front office protocol;
- Board of Directors protocol;
- operational meeting protocol;
- coordination protocol;
- outgoing correspondence;
- incoming correspondence.

As we can see, the document types mentioned easily and naturally form a so-called ISA-hierarchy where the most general type (or class) may be named *Document* and the second-level ones fall into 5 basic groups:
- contract;
- directive;
- protocol;
- correspondence;
- other.

### 5.2. Roles
Note the way to handle exceptions - the so-called *Other Document* class.

Now let us consider roles. They fall out into the following classes:
- user;
- system administrator;
- security administrator;
- guest;
- other.

Again, as in case with documents, a natural approach seems to build *Role* class hierarchy. (Let us assume *Role* has two descendant classes: *User* and *Other* of which we won't consider *Other* further here.)

Two basic classes of *Role* which conribute essentially to document processing are *User* and *Administrator*, where *Administrator* instances have rights to directly operate with rights of *User* instances, passwords, etc. (*Security_adm*) and also deal with document store directly, create and modify system help (*System_adm*) while *User* instances are not granted these rights. *User* instances, however, have rights to create and modify documents and document profiles. At this *User* class can be roughly divided into *Boss* and *Secretary* classes. Basic difference between a *Boss* instance and a *Secretary* instances that the former one has a right to sign a document while the latter one lacks this right. *Boss* and *Secretary* roles have much in common, varying, however, in access levels that depend in their turn on corporation personnel structure.

For example, access level of a Department Chief can not exceed that of a Corporation Vice-President.

**5.3. Routes**
Some document routes are pre-determined. However, it is very hard or next to impossible to explicitly define all the possible point-to-point routes including various exceptions like *Other Document* or *Guest User*. That is why the so-called hybrid routing strategy is a posiible solution in this complicated case. It basically means that some routes should be explicitly defined (as point-to-point or in a different way) while others should be whithin a pre-defined range or spectrum of possible routes.

Though there is a quite a number of theoretical foundations to make a model for an arbitrary system (taking into account the properties of the system to be designed) several rigorous, really suitable and natural approaches could be recommended.

First of all we need to control document flow. A formalism dealing with this is data flow diagrams (DFD) same as a methodology for CASE technologies. However, this formalism has a more formal and rigorous approach under which is the so-called *lattice theory* [1]. According to this theory information in general can be presented in a form of a lattice representing directions of possible information flows.

Another task is to make document structuring. Archival and query effectiveness depends on whether this task is performed in a right way or not. Object approach suggests the "right " (meaning natural and rigorous) way to be document hierarchy. One of the solid ways to implement such hierarchy is a concept of *ISA hierarchy* suggested by Roussopulos [4]. This is a basic uniform approach which holds true for concepts, objects and classes.

**6. Implementation Suggestions and Interface Discussion**
The above mentioned system examples reveal that to construct a true automated document management and control system a complex solution is needed. At least two solutions are possible:

- a program complex where document management, document control and document mailing systems are integrated;
- an integral all-in-one document management, control and mailing system.

To get a more universal and true versatile solution the second approach is better. However, for both of them a close integration of recent advances in database systems and CASE technologies is desired.

A possible (and a practically approved) combination could be Sybase Powerbuilder as a RAD tool and ER-Win as a means to manage (and to control) data flows.

Complex objects processing, differentiated righs data access, routing and a large amount of related problems arise when designing and implementing any office sys tem which is versatile enough. Hierarchical approach to data proposed in this paper becomes even more powerful and popular today with Windows '95 era advent. A whole number of rapid application development tools appeared recently which support hierarchy and trees. The leader in the list is definitely Sybase PowerBuilder 5.0 [7, 11-14].

PowerBuilder 5.0 differs essentially compared to previous 4.x versions [11,13,14]. New system object classes such as *TreeView* representing tree objects and *ListView* representing list objects appeared. The tree is implemented as follows. *TreeViewItem* element of *TreeView* object is characterised by its tree level (*TreeViewItem.level*), identifier (*handle*), data field (*TreeViewItem.data*), connected to the element, two icons corresponding to active and desactivated object states, and a large number of other parameters defining the element status (*child, expanded, collapse* etc.).

Identifier is central to *TreeViewItem* object. Identifier is a number which is defined by counting elements starting from level 1 up to the current one.Root element identifier equals 1. Zero value is root parent identifier upon root element creation. Create, delete and modify operations are valid over *TreeViewItem* element. There are no sublevels for any element just created. All sublevels are automatically deleted after a given level element has been deleted.

*ListView* is another object class worth description. Its attributes are somewhat more simple than those of *TreeView*. Namely, each *ListView* object *ListViewItem* element is characterised by an index (*index*), data field (*ListViewItem.data*), linked with the object itself, two icons corresponding to active and desactivated object states and some others. *Index* identifies *ListItem* element indicating the element position in the list.

Using both classes in the implementetion provides a user-friendly, transparent and compact interface. Tree-based implementation allows to handle a large amount complicated objects where the objects structure can be fully described by tree structure. Objects (documents, users etc.) routing is easily implemented by tree scanning along the tree branches. Both strict and flexible routing, i.e., hybrid routing is possible. Differentiating rights to document tree access limits possible access range to only limited range of tree levels, branches or selected element subtrees. User groups and document folders can be naturally implemented by trees giving transparent and effective security model resembling well known Netware Directory Service (NDS) on which Novell NetWare security system is based.

Each document profile implemented in a form of tree has a *Security_type* field with possible values (within a very basic approach) from a domain of three elements: {*Private*, *Public*, *Confidential*}. *Public* security value means each and every workgroup member has certain access type to the document or a folder depending on the document profile, routes and roles limitations. *Private* security value means only corresponding workgroup members have certain access type to the document or a folder. *Confidential* security means only authorised users explicitly named in a certain access list has certain access type to the document or a folder depending on the document profile, routes and roles limitations.

Since system efficiency and features are dependent upon documents and/or folders, workgroups, users and/or roles tree structure, tree structure choice is the key to implementation success. Using proven theories and integrating hottest approaches and tools to implement the system in question asssures state-of-the-art features and interface of the system implemented, producing an efiicient and secure product in the long run.

Queries can be quickly and effectively implemented using an attractive-looking patented innovation Sybase suggests which is DataWindow. DataWindow class features has been upgraded and significantly enriched in PowerBuilder 5.0 versus previous versions.
Another option is the so-called DataStore object class which is a powerful tool for handling dynamic trees. DataStore may be the same as that of DataWindow. SQL Cursors mechanism is another flexible and efficient way to handle DataStore objects.

### 7. A system example
Recent studies [5] show that no tools existing at present are capable of handling linear, table and hierarchical structured documents.

The system suggested is supposed to solve the above problem and is also capable to effectively store and handle unstructured data. Lotus Notes 4.0 Server choice allows to solve the latter problem since it is the primary objective of the system. The system benefits following features:
- multimedia data access (text, images audio and video);
- multiuser access to semi-structured data;
- multiuser client access to mail;

- advanced replication mechanism featuring the so-called server passthrough which enables reliable simulltaneous multiple databases access on multiple servers;
- GUI forms generator;
- e-mail front-end;
- tools and templates to create databases;
- formula-based scripting language;
- open API set;
- electronic signatures;
- security and access control lists;
- advanced database management tools.

However, even a product based on current groupware industry leader inherits some shortcomings of the environment it has been developed under:
- not very good structured data handling;
- multiuser updates with high integrity level are somewhat slow and inefficient;
- query intensive applications area inefficient to some extent.

A possible solution for the majority of the above mentioned problems seems to come from such capability of Lotus Notes 4.0 that has been absent in previous versions. This feature is the ability of third party scripts embedding into LotusScript. PowerSoft PowerBuilder, an object-oriented SQL-based RAD tool for four years consistently being the market leader in its class [12]. That is why authors suggest integrating PowerBuilder script into the groupware concerned. Rich object primitives collection of PowerBuilder features specific classes to represent tree and list structures. It allowed authors to build a uniform Windows'95 styled user-friendly and transparent interface on which meta-SQL query design interface can easily be implemented which allows users to manipulate with real object domains rather than table fields including primary keys and other system level information being redundant to users.

PowerBuilder also features the so-called DataWindows - a flexible and dynamically adjusting tool for queries and reports generation.

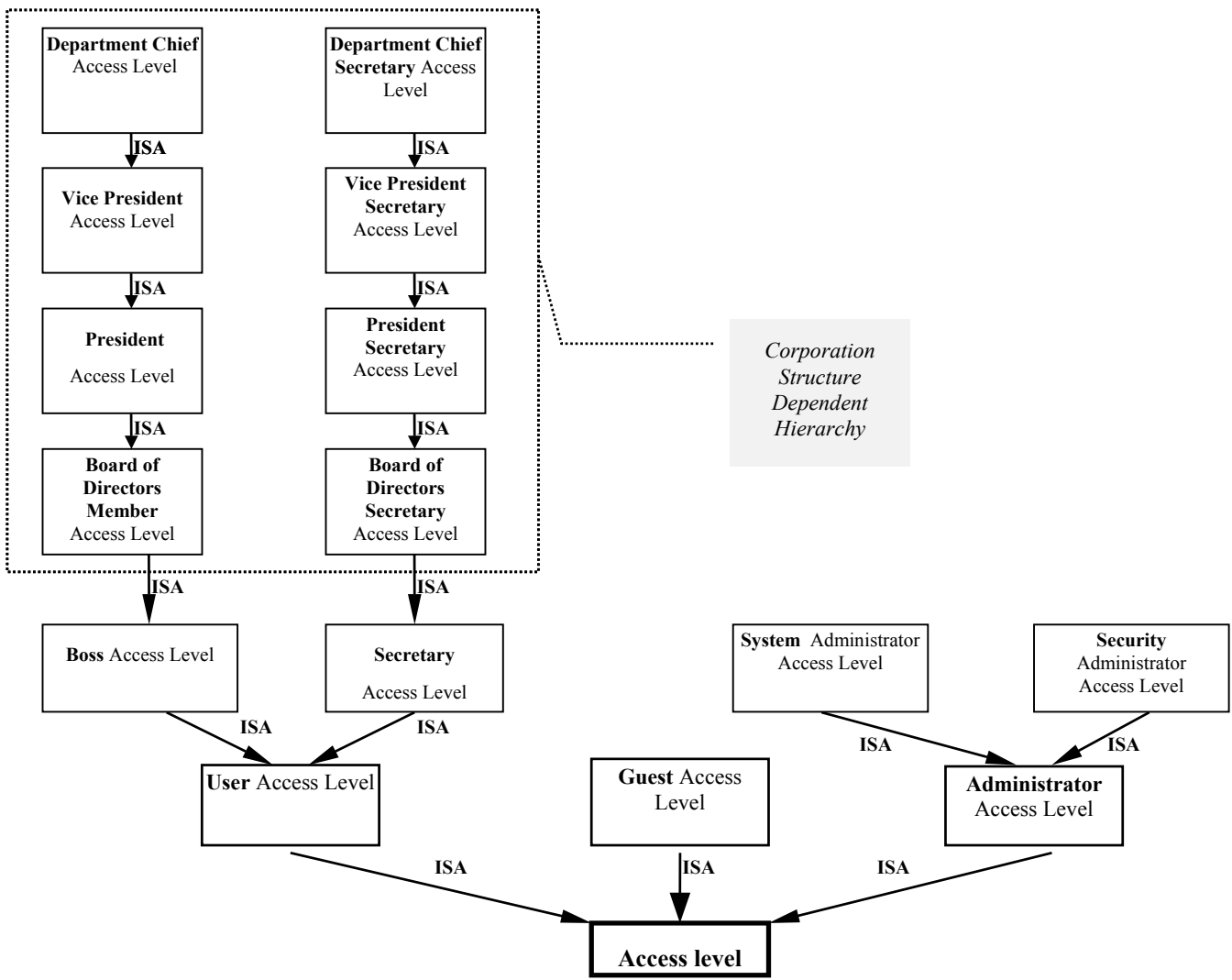

Figure1: Access Levels ISA Hierarchy

The approach concerned gives the following benefits:
- scalability;
- OS-independence;
- multi-format document handling;
- SQL databases (using popular client/server tools and front-ends) support;
- multimedia data access;
- both structured and unstructured data handling;
- Internet and Intranets support;
- flexible and dynamically adjusting queries and reports generation;
- easy to implement and efficient security system enhancements;
- user-friendly and transparent user interface.

To give a more detailed idea of what the system is like a couple of system design stage illustrations is given below.

Access level ISA hierarchy example is given in Figure1.

Document ISA hierarchy example is presented in Figure2.

based, client/server technology oriented groupware system desidn and implementation is presented. Industry trends and perspectives are given.

## 9. Conclusion and Perspectives

Groupware industry is facing new challenges on many fronts-from SQL database vendors including Sybase with their new data warehousing facilities, from the exploding Internet with its document publishing and conferencing facilities, and from CORBA Object Request Brokers. Possible trends in the groupware industryas authors can see can be summarised as follows:
- Groupware vendors are embracing the Web. The Web, with its open document standards-including document browsers, firewalls, the pervasive HTML/HTTP publishing standards, and the SMTP/MIME e-mail backbone-is the most formidable competitor groupware vendors are currently facing. Groupware vendors are in the best position to provide industrial strength technology for the Web-including scalable
- document databases, mission-critical mail backbones, security, server-to-server document

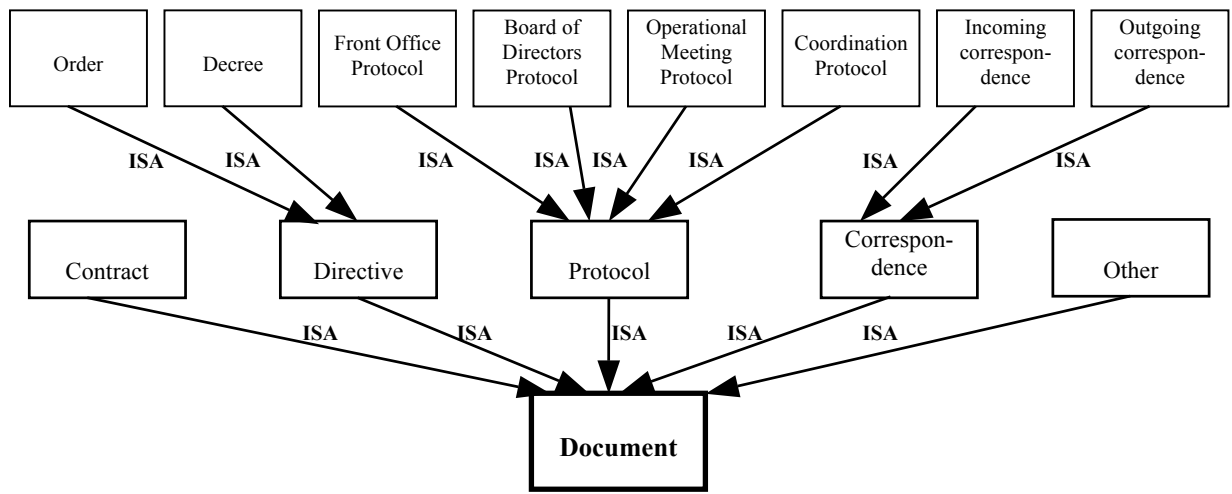

Figure 2: Document Classes ISA Hierarchy

## 8. Results

This paper presents existing approaches to groupware technologies overview. Products that are currently in the market are compared.. Lattice of flow diagram is used to model workflow control. Hierarchies in roles, routes, access control lists Groupware and client/server RAD leading products integration is proposed. The integration considered is based on solid theoretical computer science foundation. The integrated system based on the approach concerned benefits even more flexibility, user-friendly and transparentinterface and security than the components it is built upon. A scalable, robust, mission-critical, OS-independent, SQL-replications support for mobile users, workflow, and system management. Without exception, all the major groupware vendors have decided not to fight the Web, instead, they win join it. They are now in the process of recreating their groupware offerings on top of open Web standards.
- Groupware is now mission-critical ready. Lotus Notes 4.0 is the first product to provide a truly robust, scalable, and OS-independent groupware infrastructure. The product now includes important backbone features such as dynamic routing, replication, pass-through servers, thread-based background routing, security and global system management. Notes 4.0 has set a new bar for the robust features you should expect from a groupware server. Probably, Lotus competitors will come up with products that are just as robust and mission-critical. So, groupware has finally moved from the experimental stage to the production stage.

- Groupware is expanding into new areas. If they succeed in subsuming the Internet, the groupware vendors will be in a good position to subsume other forms of client/server-including database and TP Monitors. Groupware is also moving into new areas such as telephony. Finally groupware's workflow and agent-based mail systems are strong candidates for managing business processes across an enterprise.
- Groupware industry discovered that it takes great tools to win the hearts of client/server software developers. One can now create groupware applications that incorporate data from both document stores and SQL databases using popular client/server tools-including Delphi, Visual Basic, PowerBuilder, SQL Windows, and VisualAge. However, we do not know of any 3-tier client/server groupware tools - for example, a tool that lets you create workflows in the middle tier.

Groupware has come a long way The field also suffers from being a monolithic technology that is not well-suited for dealing with distributed components. In other words, groupware lacks the component middleware foundation of a distributed object bus, such as CORBA. However, groupware technology is here today and you can use it to create some very exciting client/server applications - especially on the Internet and Intranets.